\documentclass[showpacs,preprintnumbers,10pt,twocolumn]{revtex4}%
\usepackage{amssymb}
\usepackage{amsfonts}
\usepackage{amsmath}
\usepackage{graphicx}
\usepackage{dcolumn}
\usepackage{bm}
\usepackage{revsymb}%
\begin{document}
\title{Exact universal excitation waveform for optimal enhancement of directed
ratchet transport}
\author{Ricardo Chac\'{o}n$^{1}$ and Pedro J. Mart\'{\i}nez$^{2}$}
\affiliation{$^{1}$Departamento de F\'{\i}sica Aplicada, E. I. I., Universidad de
Extremadura, Apartado Postal 382, E-06006 Badajoz, Spain and Instituto de
Computaci\'{o}n Cient\'{\i}fica Avanzada (ICCAEx), Universidad de Extremadura,
E-06006 Badajoz, Spain.}
\affiliation{$^{2}$Departamento de F\'{\i}sica Aplicada, E.I.N.A., Universidad de Zaragoza,
E-50018 Zaragoza, Spain and Instituto de Ciencia de Materiales de Arag\'{o}n,
CSIC-Universidad de Zaragoza, E-50009 Zaragoza, Spain.}
\date{\today}

\begin{abstract}
The aim of the present paper is to show the existence and properties of an
\textit{exact} universal excitation waveform for optimal enhancement of
directed ratchet transport (in the sense of the average velocity). This is
deduced from the criticality scenario giving rise to ratchet universality, and
confirmed by numerical experiments in the context of a driven overdamped
Brownian particle subjected to a vibrating periodic potential. While the
universality scenario holds regardless of the waveform of the periodic
vibratory excitations involved, it is shown that the enhancement of directed
ratchet transport is optimal when the impulse transmitted by those excitations
(time integral over a half-period) is maximum. Additionally, the existence of
a \textit{frequency-dependent} optimal value of the relative amplitude of the
two excitations involved is illustrated in the simple case of harmonic excitations.

\end{abstract}

\pacs{}
\maketitle

The possibility of generating directed transport from a fluctuating
environment without any net external force, the ratchet effect [1-3], has been
a major research topic in distinct areas of science over the last few decades.
The reasons are its potential applications for understanding such systems as
molecular motors [4], protein translocation processes [5], and coupled
Josephson junctions [6], and its wide range of potential technological
applications including the design of micro- and nano-devices suitable for
on-chip implementation. Directed ratchet transport (DRT) is now understood
qualitatively to be a result of the interplay of nonlinearity, symmetry
breaking [7], and non-equilibrium fluctuations including temporal noise [2],
spatial disorder [8], and quenched temporal disorder [9]. But only recently
have several fundamental aspects begun to be elucidated, including current
reversals [10] and the quantitative dependence of DRT strength on the system's
parameters [11]. At first sight, this aspect of controllability should be
easier to investigate in non-chaotic physical contexts such as those of
certain extremely small systems, including many nanoscale devices and systems
occurring in biological and liquid environments, in which DRT is often
suitably described by overdamped ratchets [2,12-14]. Thus, the interplay
between thermal noise and symmetry breaking in the DRT of a Brownian particle
moving on a periodic substrate subjected to a homogeneous temporal biharmonic
excitation has been explained quantitatively in coherence with the
degree-of-symmetry-breaking (DSB) mechanism [15], as predicted by the theory
of ratchet universality (RU) [16]. For deterministic ratchets subjected to
\textit{biharmonic} forces, it has been shown [16] that there exists a
universal force waveform which optimally enhances directed transport by
symmetry breaking. Specifically, such a particular waveform has been shown to
be unique for both temporal and spatial biharmonic forces. This universal
waveform is a direct consequence of the DSB mechanism: It is possible to
consider a quantitative measure of the DSB on which the strength of directed
transport by symmetry breaking must depend. This mechanism has led to the
unveiling of a criticality scenario for DRT. Indeed, it has been shown that
optimal enhancement of DRT is achieved when maximal effective (i.e.,
\textit{critical}) symmetry breaking occurs, which is in turn a consequence of
two reshaping-induced competing effects: the increase of the DSB and the
decrease of the (normalized) maximal transmitted impulse over a half-period
($I\left[  f\right]  \equiv\left\vert \int_{T/2}f(t)dt\right\vert $ [16]),
thus implying the existence of a particular force waveform which optimally
enhances DRT. The definition of the DSB of the symmetries of a $T$-periodic
zero-mean ac force $f\left(  z\right)  $ is included here for the sake of
completeness:%
\begin{align}
D_{s}\left(  f\right)   &  \equiv\left\langle \frac{-f\left(  z+T/2\right)
}{f(z)}\right\rangle _{T}\equiv\frac{1}{T}\int_{0}^{T}\frac{-f\left(
z+T/2\right)  }{f(z)}dz,\nonumber\\
D_{+}\left(  f\right)   &  \equiv\left\langle \frac{f\left(  -z\right)
}{f\left(  z\right)  }\right\rangle _{T}\equiv\frac{1}{T}\int_{0}^{T}%
\frac{f\left(  -z\right)  }{f\left(  z\right)  }dt,\nonumber\\
D_{-}\left(  f\right)   &  \equiv-D_{+}\left(  f\right)  , \tag{1}%
\end{align}
where increasing deviation of $D_{s,+,-}\left(  f\right)  $ from 1(unbroken
shift and reversal symmetries, respectively) indicates an increase in the DSB
and $z=\left\{  t,x\right\}  $ (see [16] for additional details). Given the
existence of such a universal waveform whose \textit{biharmonic} approximation
is now known, the following fundamental questions naturally arise: What is the
\textit{exact} waveform of such a universal periodic force? What are the
geometric properties of the associated optimal ratchet potential?

We shall here deduce the existence and properties of such an exact universal
excitation waveform from the criticality scenario by providing two alternative
derivations, and explore its implications in the case of a driven Brownian
particle moving in a back-and-forth travelling periodic potential [2]
described by the overdamped model%
\begin{equation}
\overset{.}{x}+\sin\left[  x-\gamma f\left(  t\right)  \right]  =\sqrt{\sigma
}\xi\left(  t\right)  +\gamma g\left(  t\right)  , \tag{2}%
\end{equation}
where $f(t),g(t)$ are temporal excitations with zero mean, $f(t)$ is
$T$-periodic, $\gamma$ is an amplitude factor, $\xi\left(  t\right)  $ is a
Gaussian white noise with zero mean and $\left\langle \xi\left(  t\right)
\xi\left(  t+s\right)  \right\rangle =\delta\left(  s\right)  $, and
$\sigma=2k_{b}T^{\prime}$ with $k_{b}$ and $T^{\prime}$ being the Boltzmann
constant and temperature, respectively. Note that Eq.~(2) is equivalent to%
\begin{align}
\overset{.}{z}+\sin z  &  =\sqrt{\sigma}\xi\left(  t\right)  +\gamma F\left(
t\right)  ,\nonumber\\
F\left(  t\right)   &  \equiv g\left(  t\right)  -\overset{.}{f}\left(
t\right)  , \tag{3}%
\end{align}
where $z(t)\equiv x(t)-\gamma f\left(  t\right)  $, and $z$ and $x$ are the
particle phases relative to the vibrating potential frame and the laboratory
frame, respectively. Since the mean velocity on averaging over different
realizations of noise is the same in both frames, $\left\langle \left\langle
\overset{.}{x}\right\rangle \right\rangle =\left\langle \left\langle
\overset{.}{z}\right\rangle \right\rangle $, we shall consider Eq.~(3) for
convenience in our analysis. For the sake of clarity, we shall confine
ourselves to the regime where the DSB mechanism dominates over the thermal
inter-well activation mechanism [15]. Also, we shall show how RU allows the
dependence of DRT velocity on the system's parameters to be explained
quantitatively, and works effectively in two significant cases: (1) when
$F(t)$ is a truncated Fourier series of the exact universal periodic
excitation after $N\geqslant2$ terms, and (2) when $f(t)$ and $g(t)$ are
harmonic excitations. For deterministic ratchets, the effectiveness of the
theory of RU has been demonstrated in diverse physical contexts in which the
driving excitations are chosen to be biharmonic. Examples are cold atoms in
optical lattices [17], topological solitons [9], Bose-Einstein condensates
exposed to a sawtooth-like optical lattice potential [18], matter-wave
solitons [11], and one-dimensional granular chains [19].

\textit{Exact universal excitation waveform}.$-$Let us assume in this section
that the excitation's amplitude and period are fixed. The criticality scenario
giving rise to the existence of a universal excitation waveform which
optimally enhances DRT is a consequence of two competing reshaping-induced
effects: the increase in DSB and the decrease in the (normalized) maximal
transmitted impulse over a half-period [16]. This means that the greater the
impulse transmitted by a periodic excitation having its shift symmetry broken,
the lower the DSB needed to yield the same strength of DRT, and vice versa.
Since the strength of any transport (induced by symmetry breaking or not,
i.e., by non-zero-mean forces), in the sense of the mean kinetic energy per
unit of mass on averaging over different realizations of noise $\left\langle
\left\langle \overset{.}{x}^{2}\right\rangle \right\rangle /2$, depends upon
the impulse transmitted by the driving excitation (see the Appendix for a
detailed deduction), and the waveform yielding \textit{maximal} transmitted
impulse is that of a square-wave, the exact universal waveform should present
a constant positive value, $A$, over a certain range $t\in\left[
0,\tau\right]  ,0<\tau<T$, and a constant negative value, $-B$, over the
remaining range $t\in\left]  \tau,T\right]  $, i.e., it should belong to the
parameterized family of functions%
\begin{align}%
%TCIMACRO{\tciFourier}%
%BeginExpansion
\mathcal{F}%
%EndExpansion
(t)  &  \equiv\frac{2(A+B)}{\pi}\sum_{n=1}^{\infty}\frac{\sin\left(  n\pi
\tau/T\right)  }{n}\cos\left[  \frac{2n\pi}{T}\left(  t-\tau/2\right)  \right]
\nonumber\\
&  =\frac{(A+B)}{\pi}\sum_{n=1}^{\infty}\left[  a_{n}\left(  \tau\right)
\cos\left(  n\omega t\right)  +b_{n}\left(  \tau\right)  \sin\left(  n\omega
t\right)  \right]  ,\nonumber\\
a_{n}\left(  \tau\right)   &  \equiv\frac{\sin\left(  n\omega\tau\right)  }%
{n},\ b_{n}\left(  \tau\right)  \equiv\frac{1-\cos\left(  n\omega\tau\right)
}{n}, \tag{4}%
\end{align}
where $\omega\equiv2\pi/T$. Clearly, the constraints $A\neq B$ and $\tau\neq
T/2$ are \textit{necessary }conditions to satisfy two requirements: the
breaking of the shift symmetry and the zero-mean property of the exact
universal excitation $f_{u}(t)$. This further requirement implies the
relationship
\begin{equation}
\tau=T/(1+A/B), \tag{5}%
\end{equation}
i.e., one only has to obtain the suitable value of either the asymmetry
parameter $\tau$ or $A/B$ that makes the DSB maximally effective, thus
providing the exact universal excitation waveform.

The suitable value of $\tau$ can be calculated from the observation that the
exact universal excitation waveform cannot be independent of the biharmonic
universal excitation waveform due to the \textit{unique} character of both
waveforms. This is due to the latter should inevitably be contained in the
Fourier series of the former in the form of an infinity of harmonic pairs
whose frequencies are one double the other while having the same waveform than
that of the biharmonic universal excitation. Indeed, the biharmonic universal
excitation is equivalently described by the expressions [16]%
\begin{align}
f_{\sin,\sin,\pm}(t) &  \equiv\varepsilon\left[  \sin\left(  \omega t\right)
\pm\frac{1}{2}\sin\left(  2\omega t\right)  \right]  ,\nonumber\\
f_{\cos,\sin,\pm}(t) &  \equiv\varepsilon\left[  \cos\left(  \omega t\right)
\pm\frac{1}{2}\sin\left(  2\omega t\right)  \right]  ,\tag{6}%
\end{align}
which satisfy the symmetries%
\begin{align}
f_{\sin,\sin,+}(t+T/2) &  =-f_{\sin,\sin,-}(t),\nonumber\\
f_{\cos,\sin,+}(t+T/2) &  =-f_{\cos,\sin,-}(t),\nonumber\\
f_{\sin,\sin,\pm}(t+T/4) &  =f_{\cos,\sin,\mp}(t)\tag{7}%
\end{align}
(see Fig. 1, top panel). From the Fourier series of $%
%TCIMACRO{\tciFourier}%
%BeginExpansion
\mathcal{F}%
%EndExpansion
(t)$ [Eq. (4)], one has four harmonic pairs having frequencies $\omega$ and
$2\omega$ in each pair:%
\begin{align}
&  b_{1}\left(  \tau\right)  \sin\left(  \omega t\right)  +b_{2}\left(
\tau\right)  \sin\left(  2\omega t\right)  ,\tag{8a}\\
&  a_{1}\left(  \tau\right)  \cos\left(  \omega t\right)  +b_{2}\left(
\tau\right)  \sin\left(  2\omega t\right)  ,\tag{8b}\\
&  b_{1}\left(  \tau\right)  \sin\left(  \omega t\right)  +a_{2}\left(
\tau\right)  \cos\left(  2\omega t\right)  ,\tag{8c}\\
&  a_{1}\left(  \tau\right)  \cos\left(  \omega t\right)  +a_{2}\left(
\tau\right)  \cos\left(  2\omega t\right)  .\tag{8d}%
\end{align}
We see that the waveforms of the biharmonic expressions (8c) and (8d) do not
correspond to that of the biharmonic universal excitation (cf. Eq. (6)), the
biharmonic expression (8b) with $a_{1}\left(  \tau\right)  =\pm2b_{2}\left(
\tau\right)  $ does but presents a phase difference of $T/4$ with respect to $%
%TCIMACRO{\tciFourier}%
%BeginExpansion
\mathcal{F}%
%EndExpansion
(t)$, while the biharmonic expression (8a) with $b_{1}\left(  \tau\right)
=\pm2b_{2}\left(  \tau\right)  $ does and is in phase with $%
%TCIMACRO{\tciFourier}%
%BeginExpansion
\mathcal{F}%
%EndExpansion
(t)$. Therefore, the compatibility between the exact universal excitation
waveform and the biharmonic universal excitation \textit{requires} that
$\left\vert b_{1}\left(  \tau\right)  /b_{2}\left(  \tau\right)  \right\vert
=2$, i.e.,%
\begin{equation}
1-\cos\left(  \omega\tau\right)  =\pm\frac{1-\cos\left(  2\omega\tau\right)
}{2}.\tag{9}%
\end{equation}
After defining $z\equiv\cos\left(  \omega\tau\right)  $, Eq. (9) can be put
into the form $2z^{2}-z-1=0$, $2z^{2}-z-3=0$, for the signs $+,-$,
respectively. The solutions $z=-3/2$ and $z=1$ of the latter algebraic
equation lack mathematical sense ($z=-3/2<-1$) and physical sense
($z=1\Rightarrow A/B=0$, cf. Eq. (5)), respectively. The solutions of the
former algebraic equation are $z=\left\{  1,-1/2\right\}  $. For the only
meaningful solution, $z=-1/2$, one has%
\begin{equation}
\omega\tau\equiv\frac{2\pi\tau}{T}=\left\{
\begin{array}
[c]{cc}%
4\pi/3\Rightarrow & \tau=\frac{2}{3}T\\
2\pi/3\Rightarrow & \tau=\frac{1}{3}T
\end{array}
\right\}  .\tag{10}%
\end{equation}
Thus, after using Eq. (5), one finally obtains the conditions $\tau
=2T/3\Leftrightarrow A/B=1/2$ and $\tau=T/3\Leftrightarrow A/B=2$ for the
cases $A<B$ and $A>B$, respectively. Therefore, the values $\tau
=2T/3,\ A/B=1/2$ (or equivalently $\tau=T/3,\ A/B=2$) fix the exact universal
waveform of the excitation $f_{u}(t)$ which yields DRT having the same
strength but opposite direction in the cases $A/B=1/2$ and $A/B=2$. It is
worth noting that, for these two values of $\tau$, $\tau_{u}\equiv\left\{
T/3,2T/3\right\}  $, the Fourier coefficients of the exact universal
excitation satisfy the properties (cf. Eq. (4))%
\begin{align}
\left(  n+3\right)  b_{n+3}(\tau &  =\tau_{u})=nb_{n}(\tau=\tau_{u}%
),\tag{11a}\\
\left(  n+3\right)  a_{n+3}(\tau &  =\tau_{u})=na_{n}(\tau=\tau_{u}%
),\tag{11b}\\
a_{3n}(\tau &  =\tau_{u})=b_{3n}(\tau=\tau_{u})=0,\tag{11c}\\
b_{n}(\tau &  =\tau_{u})=2b_{2n}(\tau=\tau_{u}),\tag{11d}\\
a_{n}(\tau &  =\tau_{u})=2a_{2n}(\tau=\tau_{u}),\tag{11e}\\
b_{n}(\tau &  =\tau_{u})=\sqrt{3}\left(  -1\right)  ^{n}a_{n}(\tau=\tau
_{u}).\tag{11f}%
\end{align}
Properties (11a) and (11b) indicate a subtle periodicity of the coefficients,
while property (11c) makes explicit the periodic absence of an infinity of
coefficients. Remarkably, properties (11d) and (11e) indicate that the
harmonic pairs of the types $b_{n}\left(  \tau\right)  \sin\left(  n\omega
t\right)  +b_{2n}\left(  \tau\right)  \sin\left(  2n\omega t\right)  $ and
$a_{n}\left(  \tau\right)  \cos\left(  n\omega t\right)  +a_{2n}\left(
\tau\right)  \cos\left(  2n\omega t\right)  $, respectively, \textit{also}
satisfy the requirement of the biharmonic universal excitation regarding the
relative amplitude of the two harmonics of each pair. Notice that property
(11d) also shows that \textit{the biharmonic universal excitation waveform is
present in an infinite series of harmonic pairs}. Moreover, property (11f)
together with properties (11a), (11b), and (11c) suggest that the complete
Fourier series of the exact universal excitation $f_{u}(t)$ can be understood
as the sum of two complementary series: a series consisting only of sine terms
containing all the ratcheting effect, and another series consisting only of
cosine terms yielding the maximization of the transmitted impulse. Indeed, for
the case $A/B=1/2$ for instance, one has
\begin{align}
\frac{2\pi f_{u}(t)}{3\sqrt{3}A} &  \equiv-\cos(\omega t)+\frac{1}{2}%
\cos(2\omega t)-\frac{1}{4}\cos\left(  4\omega t\right)  \nonumber\\
&  +\frac{1}{5}\cos\left(  5\omega t\right)  -\frac{1}{7}\cos\left(  7\omega
t\right)  +\frac{1}{8}\cos\left(  8\omega t\right)  -...\nonumber\\
&  +\sqrt{3}\sin(\omega t)+\frac{\sqrt{3}}{2}\sin(2\omega t)+\frac{\sqrt{3}%
}{4}\sin\left(  4\omega t\right)  \nonumber\\
&  +\frac{\sqrt{3}}{5}\sin\left(  5\omega t\right)  +\frac{\sqrt{3}}{7}%
\sin\left(  7\omega t\right)  +\frac{\sqrt{3}}{8}\sin\left(  8\omega t\right)
\nonumber\\
&  +...\equiv C_{u}(t)+S_{u}(t),\tag{12}%
\end{align}
where $C_{u}(t),S_{u}(t)$ represent the aforementioned complementary series,
while $C_{u,N}(t),S_{u,N}(t)$ denote the corresponding truncated series after
$N$ terms, respectively (see Fig. 1, middle and bottom panels).
\begin{figure}
\includegraphics[width=0.45\textwidth]{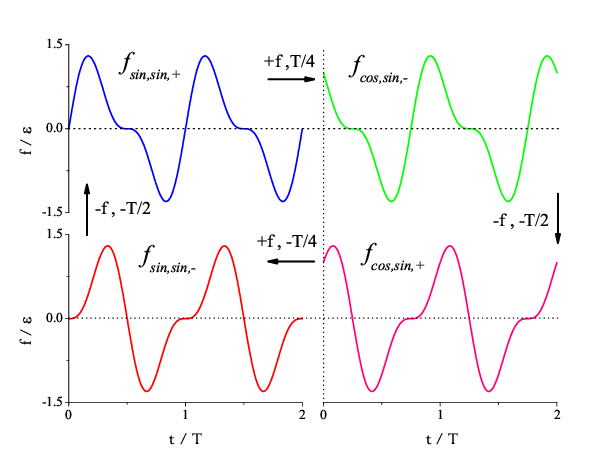}
\includegraphics[width=0.45\textwidth]{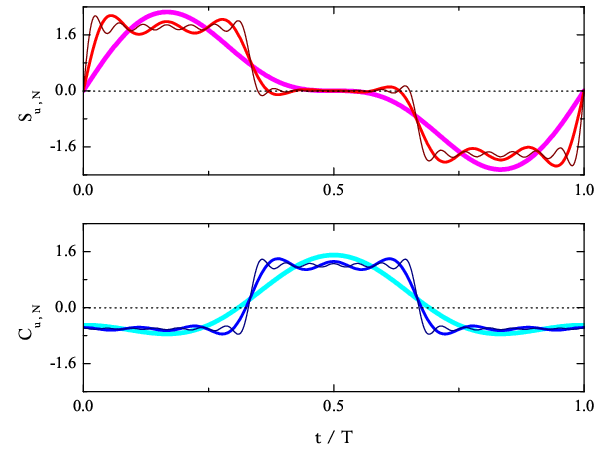}
\includegraphics[width=0.45\textwidth]{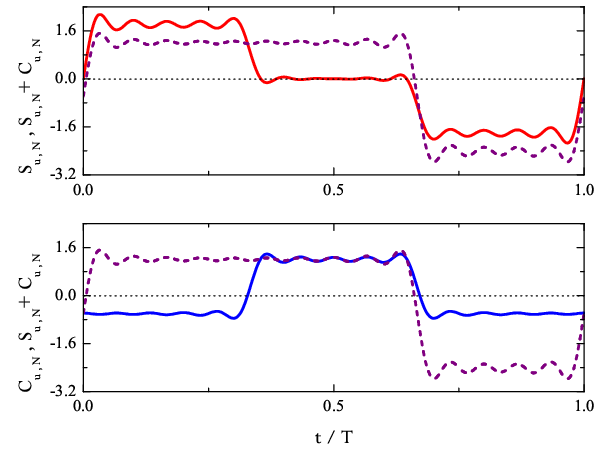}
\caption{Top: Functions $f_{\sin,\sin,\pm}(t)$ and $f_{\cos,\sin,\pm}(t)$ [cf.
Eq. (6)] representing the biharmonic universal excitation vs $t$. The
horizontal and vertical arrows indicate the symmetries that relate the
different representations [cf. Eq. (7)]. Middle: Truncations of the series
$S_{u}(t)$ and $C_{u}(t)$ [cf. Eq. (12)] after $N=2,6,14$ terms vs $t$ (solid
curves of respectively decreasing thickness), $S_{u,N}(t)$ (upper panel) and
$C_{u,N}(t)$, respectively. Bottom: Functions $S_{u,N=10}(t)$, $C_{u,N=10}%
(t)+S_{u,N=10}(t)$ (upper panel, solid and dashed lines, respectively) and
$C_{u,N=10}(t)$, $S_{u,N=10}(t),C_{u,N=10}(t)+S_{u,N=10}(t)$ (solid and dashed
lines, respectively) vs $t$.}
\label{fig1}
\end{figure}

Alternatively, the suitable value of $A/B$ can be calculated from the
quantifier of the DSB associated with the shift symmetry of $f_{u}(t)$,
$D_{s}(f_{u})$ [cf. Eq. (1)]. To this end, we properly require that the
(positive and negative) amplitudes of $%
%TCIMACRO{\tciFourier}%
%BeginExpansion
\mathcal{F}%
%EndExpansion
(t)$ and a suitable (symmetry-breaking-inducing) biharmonic excitation, for
example $f_{bh}(t)=\gamma\left[  \eta\sin\left(  \omega t\right)  +\left(
1-\eta\right)  \sin\left(  2\omega t+\varphi\right)  \right]  $ with
$\gamma>0,\eta\in\left[  0,1\right]  ,\varphi=\varphi_{opt}\equiv\pi/2$ [16],
should be the same, i.e., $A=\max_{t}f_{bh}(t;\varphi_{opt}\equiv\pi/2)$,
$B=-\min_{t}f_{bh}\left(  t;\varphi_{opt}\equiv\pi/2\right)  $. One thus
obtains straightforwardly%
\begin{align}
D_{s}\left(  f_{u}\right)   &  \equiv\frac{1}{T}\int_{0}^{T}\frac
{-f_{u}\left(  t+T/2\right)  }{f_{u}(t)}dt=\frac{A}{B}=\frac{T-\tau}{\tau
}\nonumber\\
&  =\left\{
\begin{array}
[c]{cc}%
1-\eta+\frac{\eta^{2}}{8\left(  1-\eta\right)  }, & \eta\leqslant\frac{4}{5}\\
2\eta-1, & \eta\geqslant\frac{4}{5}%
\end{array}
\right\}  ,\tag{13}%
\end{align}
with $A<B$ (and hence $T/2<\tau<T$), and where an increase in the deviation of
$D_{s}\left(  f_{u}\right)  $ from 1 (unbroken symmetry) indicates an increase
in the DSB. One finds that $D_{s}\left(  f_{u}\right)  $ has the value
$D_{s}\left(  f_{u}\right)  \mid_{\eta=0,1}=1$, and presents, as a function of
$\eta$, a single extremum at $\eta=2/3$ (see Fig.~2, top panel), and hence the
DSB is maximum when $A/B=1/2,\tau=2T/3$ [cf. Eqs.~(5) and (13)]. As expected
from a symmetry analysis, we obtained the same behaviour when using any other
alternative form for $f_{bh}(t)$ together with the corresponding suitable
values of $\varphi_{opt}$ in each case [16]. In particular, for the other
optimal value, $\varphi_{opt}\equiv3\pi/2$ corresponding to $f_{bh}%
(t)=\gamma\left[  \eta\sin\left(  \omega t\right)  +\left(  1-\eta\right)
\sin\left(  2\omega t+\varphi\right)  \right]  $, one straightforwardly
obtains $\max_{t}f_{bh}(t;\varphi_{opt}\equiv3\pi/2)=-\min_{t}f_{bh}\left(
t;\varphi_{opt}\equiv\pi/2\right)  $, $\min_{t}f_{bh}\left(  t;\varphi
_{opt}\equiv3\pi/2\right)  =-\max_{t}f_{bh}(t;\varphi_{opt}\equiv\pi/2)$, and
$D_{s}\left(  f_{u}\right)  =B/A=\tau/\left(  T-\tau\right)  $ with $A>B$ (and
hence $0<\tau<T/2$). This value of $D_{s}\left(  f_{u}\right)  $ presents the
same dependence on $\eta$ than that corresponding to $\varphi_{opt}\equiv
\pi/2$ [Eq. (13)], and hence the DSB is maximum when $A/B=2,\tau=T/3$ and the
DRT has the same strength but opposite direction to that corresponding to
$\varphi_{opt}\equiv\pi/2$. Therefore, the values $A/B=1/2,\tau=2T/3$ (or
equivalently $A/B=2,\tau=T/3$) again fix the exact universal waveform of the
excitation $f_{u}(t)$ as well as the properties of the associated ratchet
potential $U_{u}\left(  x\right)  \equiv-\int f_{u}(x)dx$ (see Fig.~2, middle
and bottom panels).
In this regard, it is worth mentioning that the
biparametric $\left(  A,B\right)  $ family of dichotomous driving waveforms
predicted in Ref.~[20] for optimal enhancement of DRT in overdamped, adiabatic
rocking ratchets includes (without indicating that it is a special case) the
exact universal waveform of $f_{u}(t)$ for the particular choice $A/B=1/2$.
Also, the exact universal waveform was used (without indicating the reason of
its choice) in the experimental realization of a
relativistic-flux-quantum-based diode [12]. After calculating the Fourier
series of the universal excitation and potential,%
\begin{align}
f_{u}(t) &  \equiv\frac{6A}{\pi}\sum_{n=1}^{\infty}\frac{\sin\left(
2n\pi/3\right)  }{n}\cos\left[  2n\pi\left(  t/T-1/3\right)  \right]
,\tag{14}\\
U_{u}(x) &  \equiv-\frac{3A\lambda}{\pi^{2}}\sum_{n=1}^{\infty}\frac
{\sin\left(  2n\pi/3\right)  }{n^{2}}\sin\left[  2n\pi\left(  x/\lambda
-1/3\right)  \right]  ,\tag{15}%
\end{align}
where $\lambda$ is the spatial period, one obtains the geometric properties of
the universal ratchet potential per unit of amplitude and unit of spatial
period [Eq.~(15); see Fig.~2, bottom panel].
\begin{figure}[ht]
\includegraphics[width=.42\textwidth]{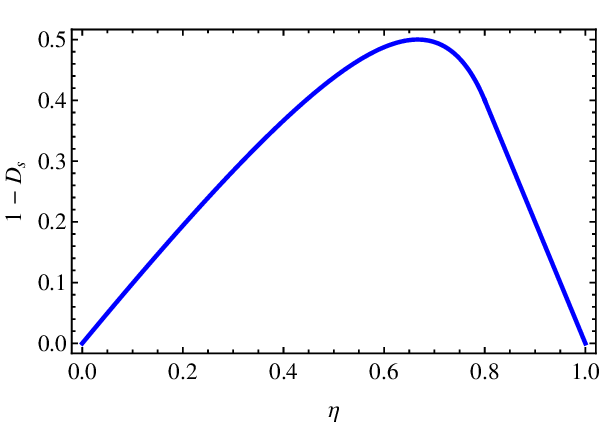}
\includegraphics[width=.42\textwidth]{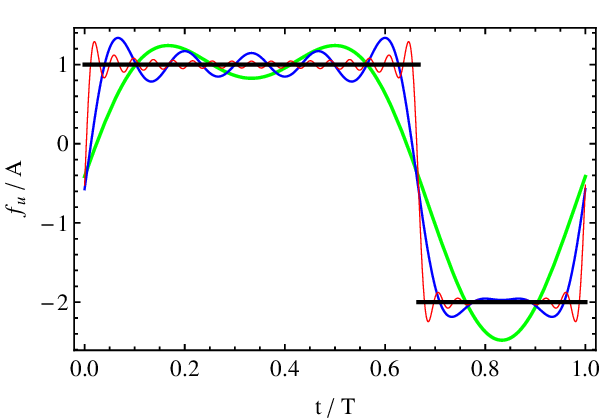}
\includegraphics[width=.42\textwidth]{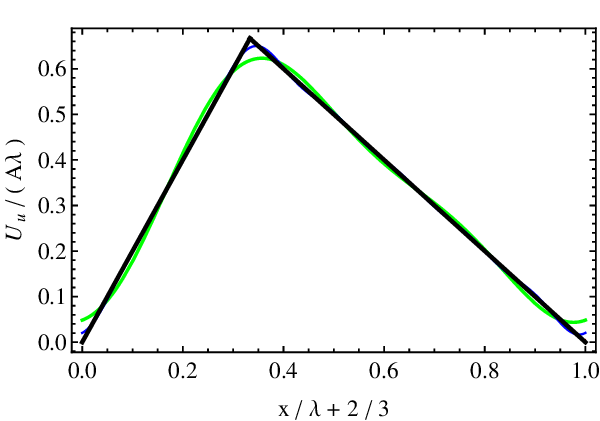}
\caption{Top: Quantifier of the DSB associated with the shift symmetry $D_{s}$
vs amplitude factor $\eta$ [cf.~Eq.~(13) and the text] for the exact universal
excitation $f_{u}(t)$ [cf.~Eq.~(14)]. Middle: Function $f_{u}(t)$ and the
truncations of its Fourier series after $N=2,7,25$ terms vs $t$ (solid curves
of respectively decreasing thickness). Bottom: Exact universal potential
$U_{u}(x)$ [cf.~Eq.~(15)] and the truncations of its Fourier series after
$N=2,7,25$ terms vs $x$ (solid curves of respectively decreasing thickness).
The values of the steep and shallow slopes are $2$ and $-1$, respectively.}
\label{fig2}
\end{figure}

 Next, we consider the case
$g(t)=0$ and $f(t)=$ $-\left(  1/A\right)  \int f_{u,N}(t)dt$ [cf.~Eq.~(15)],
i.e., $F(t)\equiv f_{u,N}(t)/A$ in Eq.~(3), with $f_{u,N}(t)$ being the
Fourier series of $f_{u}(t)$ truncated after $N$ terms [cf.~Eq.~(14)]. Our
numerical results systematically indicate an overall increase of the maximum
value of $\left\langle \left\langle \overset{.}{z}\right\rangle \right\rangle
$ with the number of terms $N$, while keeping the remaining parameters
constant. Moreover, the typical instance shown in Fig.~3 (top panel) indicates
that the average velocity (absolute value) quickly increases with $N$, and
reaches its asymptotic value for $N\sim13$. This behaviour is found to be
correlated with that of the impulse per unit of amplitude transmitted by
$f_{u,N}(t)$ over a half-period,%
\begin{align}
I_{N} &  \equiv\frac{1}{T}\int_{0}^{T/2}f_{u,N}(t)dt\nonumber\\
&  =\frac{3}{\pi^{2}}\sum_{i=1}^{N}\frac{\sin\left(  \frac{2i\pi}{3}\right)
\left[  \sin\left(  \frac{i\pi}{3}\right)  +\sin\left(  \frac{2i\pi}%
{3}\right)  \right]  }{i^{2}},\tag{16}%
\end{align}
as expected from the theory of RU [16] (see Fig.~3, bottom panel).
\begin{figure}[ht]
\includegraphics[width=.48\textwidth]{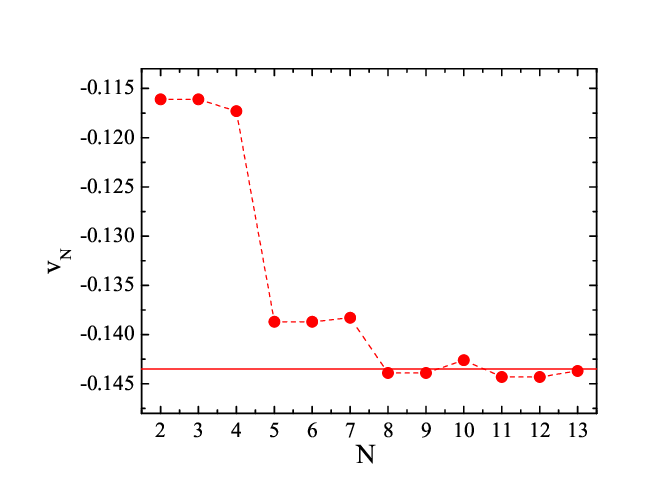}
\includegraphics[width=.48\textwidth]{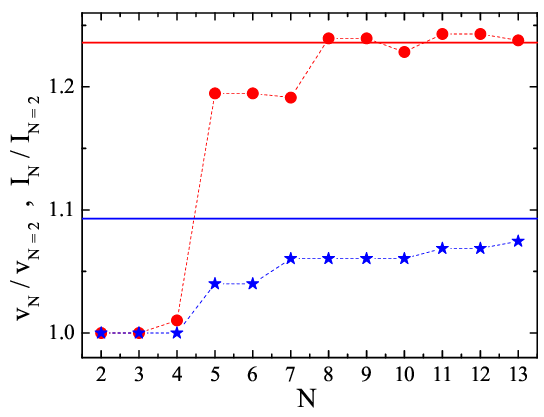}
\caption{Top: Average velocity $v_{N}\equiv\left\langle \left\langle
\overset{.}{z}\right\rangle \right\rangle $ [cf.~Eq.~(3); dots] as a function
of the number, $N$, of harmonics which are retained in the truncated Fourier
series of $f_{u}(t)$ [cf.~Eq.~(14)]. The horizontal line indicates the
asymptotic value of the average velocity corresponding to the complete series
of $f_{u}(t)$. Bottom: Normalized average velocity (dots) and normalized
impulse [cf.~Eq.~(16); stars] as functions of the number of harmonics, $N$.
The horizontal lines indicate the respective asymptotic values when
$N\rightarrow\infty$. The dashed lines connecting the symbols are solely to
guide the eye. Fixed parameters: $\gamma=8,T=4\pi,\sigma=0.8$.}
\label{fig3}
\end{figure}

\textit{Harmonic excitations}.$-$For the sake of completeness, we next explore
the standard case [2] in which the two temporal excitations involved are
harmonic: $f(t)\equiv\eta\cos\left(  \omega t\right)  $, $g(t)\equiv\left(
1-\eta\right)  \cos\left(  2\omega t+\varphi\right)  ,\omega\equiv2\pi
/T,\eta\in\left[  0,1\right]  $ in Eq.~(2), i.e.,
\begin{equation}
F(t)\equiv\eta\omega\sin\left(  \omega t\right)  +\left(  1-\eta\right)
\cos\left(  2\omega t+\varphi\right)  \tag{17}%
\end{equation}
in Eq.~(3). Leaving aside the effect of noise (an effective change of the
potential barrier which is in turn controlled by the DSB mechanism [15]), RU
predicts (for $\sigma=0$) that the optimal value of the relative amplitude
$\eta$ comes from the condition that the amplitude of $\sin\left(  \omega
t\right)  $ must be twice as large as that of $\cos\left(  2\omega
t+\varphi\right)  $ in Eq.~(3) with $F(t)$ given by Eq.~(17), and the optimal
values of the initial phase difference are $\varphi=\varphi_{opt}%
\equiv\left\{  0,\pi\right\}  $ [16]. Thus, RU predicts the existence of a
\textit{frequency-dependent} optimal value of $\eta$:%
\begin{equation}
\eta_{opt}\equiv2/\left(  2+\omega\right)  ,\tag{18}%
\end{equation}
and, equivalently, an optimal frequency for each value of $\eta$:
$\omega_{opt}\equiv2\left(  1-\eta\right)  /\eta$. Numerical simulations
confirmed this prediction over a wide range of frequencies (see Fig.~4, top
panel). 
\begin{figure}[ht]
\includegraphics[width=.48\textwidth]{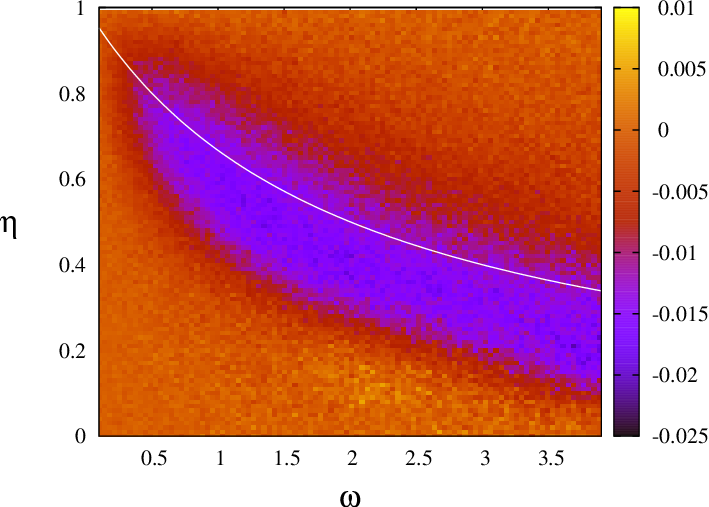}
\includegraphics[width=.48\textwidth]{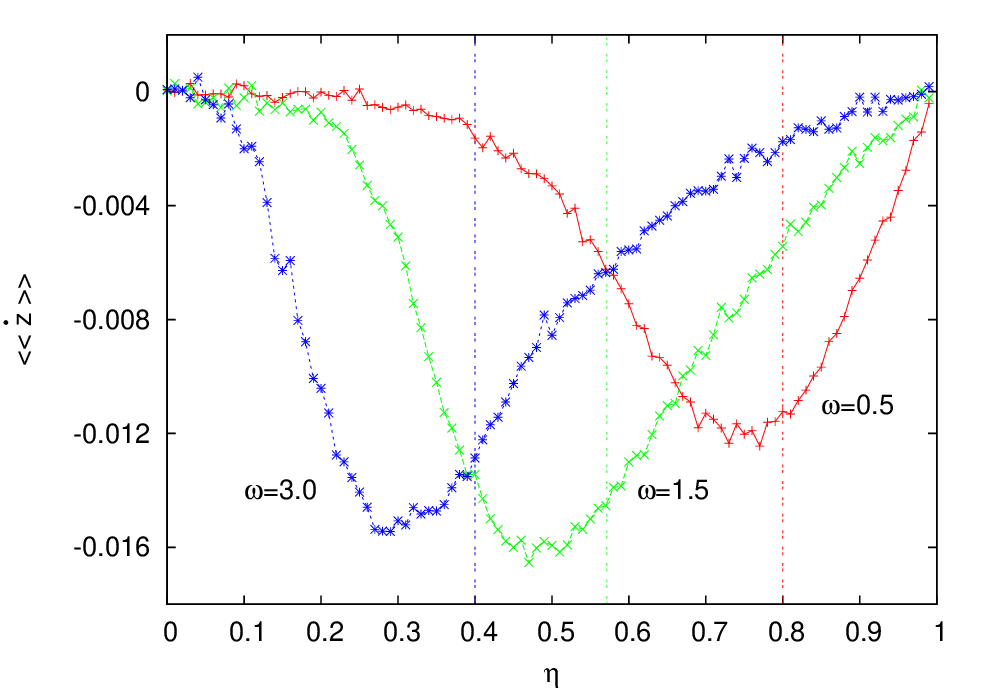}
\caption{Top: Average velocity $\left\langle \left\langle \overset{.}%
{z}\right\rangle \right\rangle $ [cf.~Eq.~(3)] vs relative amplitude $\eta$
and frequency $\omega$ for $F(t)\equiv\eta\omega\sin\left(  \omega t\right)
+\left(  1-\eta\right)  \cos\left(  2\omega t+\varphi\right)  $
[cf.~Eq.~(17)]. Also plotted is the theoretical prediction for the maximum
average velocity [cf.~Eq.~(18); solid curve]. Bottom: $\left\langle
\left\langle \overset{.}{z}\right\rangle \right\rangle $ vs $\eta$ for three
values of the frequency: $\omega=0.5,1.5,3$. The vertical dashed lines
indicate the respective predicted optimal values of $\eta$ for $\sigma=0$
[cf.~Eq.~(18)]. Fixed parameters: $\varphi=\varphi_{opt}=0,\sigma
=10,\gamma=15$.}
\label{fig4}
\end{figure}
As mentioned above, the numerical estimate of the $\eta$ value at
which the average velocity presents an extremum, $\eta_{opt}^{\sigma>0}$, is
slightly lower than the corresponding value $\eta_{opt}$ [Eq.~(18)], as
expected [15] (see Fig.~4, bottom panel). It is worth noting that the property
Eq.~(18) represents a genuine feature of the back-and-forth travelling
potential ratchet [Eq.~(2)] which is absent in the case of an overdamped
rocking ratchet [15]. Also, this finding is in sharp contrast with the
prediction coming from \textit{all} the earlier theoretical approaches [3,7,
21-23], namely, that the dependence of the average velocity should scale as
\begin{equation}
\left\langle \left\langle \overset{.}{z}\right\rangle \right\rangle \sim
\gamma^{3}\omega^{2}\eta^{2}\left(  1-\eta\right)  ,\tag{19}%
\end{equation}
which fails to explain the observed phenomenology (cf.~Fig.~4). Indeed, this
\textit{amplitudes catastrophe} comes from the assumption that the
contributions of the amplitudes of the two harmonic excitations to the average
velocity are independent. However, the existence of a universal waveform which
optimally enhances DRT implies that the two amplitudes are correlated in the
sense mentioned above. It is worth mentioning that the case where the roles
played by the harmonic excitations $\eta\cos\left(  \omega t\right)  $ and
$\left(  1-\eta\right)  \cos\left(  2\omega t+\varphi\right)  $ are
interchanged presents different optimal values of the initial phase $\varphi$
and a different dependence on the frequency of the optimal value of $\eta$,
and that numerical simulations again confirmed these predictions from RU (see
the Appendix for analytical and numerical details). To confirm the
aforementioned characteristics of the criticality scenario giving rise to the
existence of the exact universal excitation waveform, we compared the ratchet
effectiveness of the biharmonic excitation [Eq.~(17)] with that of $F(t)\equiv%
%TCIMACRO{\tciFourier}%
%BeginExpansion
\mathcal{F}%
%EndExpansion
(t)$ [cf.~Eq.~(4)] subjected to the requirement that both excitations have the
same (positive and negative) amplitudes for each value of $\eta$. Recall that
varying the amplitudes of $%
%TCIMACRO{\tciFourier}%
%BeginExpansion
\mathcal{F}%
%EndExpansion
(t)$ implies varying the asymmetry parameter $\tau$, and vice versa [cf.~Eq.
(5)], whence both $\tau$ and $A/B$ will be $\eta$-dependent so as to allow a
proper comparison of the ratchet effectiveness of these excitations. Indeed,
the results shown in Fig.~5 indicate that the DRT strength of the dichotomous
excitation is greater than that of the biharmonic excitation over (almost) the
\textit{entire} range of $\eta$ values, i.e., enhancement of DRT occurs when
the impulse transmitted is maximum regardless of the DSB of the two
excitations. One clearly sees in Fig.~5 that the greater the impulse
transmitted, the lower the DSB needed to yield the same strength of DRT, and
vice versa, as predicted from the criticality scenario. Note that the
noise-induced decrease of the optimal value of $\eta$ with respect to the
corresponding deterministic prediction, $\eta_{opt}^{\sigma=0}-\eta
_{opt}^{\sigma>0}$ [$\eta_{opt}^{\sigma=0}=0.8$; cf.~Eq.~(18)], is slightly
lower when the transmitted impulse is maximum. This provides additional
evidence for the impulse being the main quantifier of the driving
effectiveness of a periodic excitation. Additionally, robustness of the
present universality scenario is also observed when the external periodic
excitation is replaced by a chaotic excitation having the same underlying main
frequency in its Fourier spectrum (see the Appendix).
\begin{figure}[ht]
\includegraphics[width=.48\textwidth]{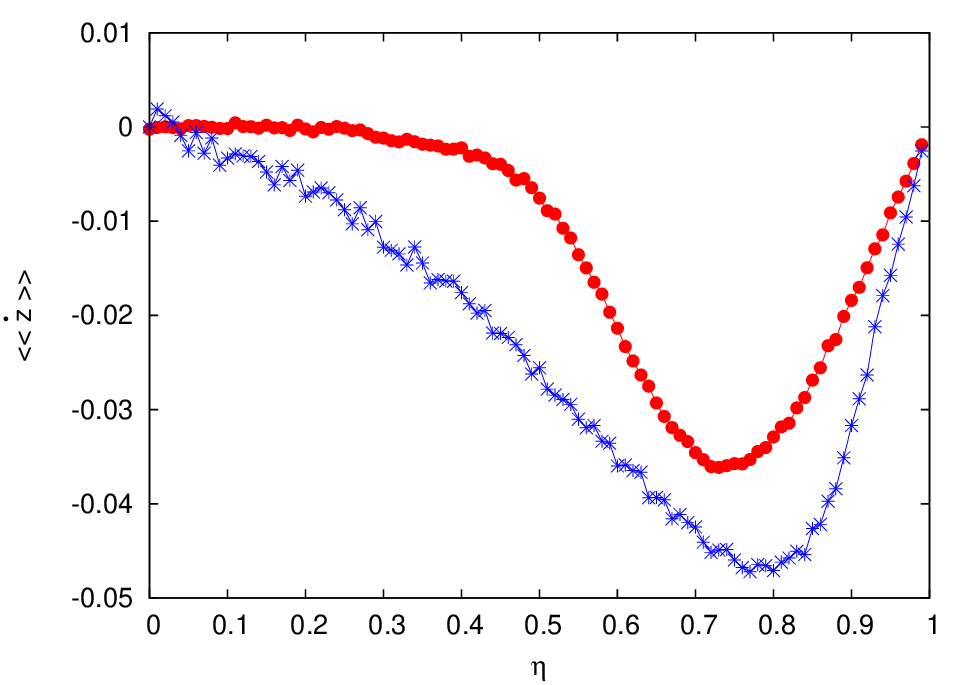}
\caption{Average velocity $\left\langle \left\langle \overset{.}{z}%
\right\rangle \right\rangle $ [cf.~Eq.~(3)] vs parameter $\eta$ (see the text)
for two choices of the excitation $F(t)$: $\eta\omega\sin\left(  \omega
t\right)  +\left(  1-\eta\right)  \cos\left(  2\omega t+\varphi\right)  $ [cf.
Eq.~(17); dots] and $%
%TCIMACRO{\tciFourier}%
%BeginExpansion
\mathcal{F}%
%EndExpansion
(t)$ [cf.~Eq.~(4); stars]. The lines connecting the symbols are solely plotted
to guide the eye. Fixed parameters: $\gamma=8,T=4\pi,\varphi=\varphi
_{opt}=0,\sigma=4$.}
\label{fig5}
\end{figure}

\textit{Conclusions.}--In summary, from the criticality scenario giving rise
to ratchet universality we have demonstrated the existence and properties of
an exact universal excitation waveform for optimal enhancement of directed
ratchet transport by providing two alternative derivations. Our numerical
experiments confirmed those findings, as well as revealed other unanticipated
properties for the standard case of harmonic excitations in the general
context of a driven overdamped Brownian particle subjected to a vibrating
periodic potential. The exact universal waveform is the \textit{simplest}
possible (a particular dichotomous waveform), and is far more efficient that
its biharmonic approximation, and the waveform of the associated optimal
ratchet potential is therefore a particular case of the simplest piecewise
waveform as is used, for instance, in a flashing ratchet. Since most models of
biological Brownian motors are compatible with a simplified description based
on the flashing ratchet, we are tempted to conjecture that the universal
optimal ratchet potential could underlie the complex biological machinery
operating at the nanoscale as a result of evolutionary processes.

R.C. acknowledges financial support from the Junta de Extremadura (JEx, Spain)
through Project No. GR18081 cofinanced by FEDER funds. P.J.M. acknowledges
financial support from the Ministerio de Econom\'{\i}a y Competitividad
(MINECO, Spain) through project FIS2017-87519 cofinanced by FEDER funds and
from the Gobierno de Arag\'{o}n (DGA, Spain) through grant E36\_17R to the
FENOL group.

\section{APPENDIX: SUPPLEMENTARY CALCULATION DETAILS AND RESULTS}

This Appendix provides details on the energy analysis, the case where the
roles of the harmonic excitations are interchanged, and the case where the
external periodic excitation is substituted by a chaotic excitation.

\subsection{Energy-based analysis}

\noindent In this subsection we deduce an analytical expression for the mean
kinetic energy per unit of mass on averaging over different realizations of
noise of a Brownian particle of mass $m$ which satisfies the general equation
of motion%
\begin{equation}
m\overset{..}{x}+\frac{dU}{dx}=-\mu\overset{.}{x}+\gamma f\left(  t\right)
+\sqrt{\sigma}\xi\left(  t\right)  , \tag{A1}%
\end{equation}
where $U(x)$ is a potential subject to a lower bound (i.e., $\exists
\ \alpha\in%
%TCIMACRO{\U{211d} }%
%BeginExpansion
\mathbb{R}
%EndExpansion
\ /\ U(x)\geqslant\alpha\ \forall x$), $f(t)$ is a unit-amplitude $T$-periodic
function with zero mean, $\xi\left(  t\right)  $ is a Gaussian white noise of
zero mean and $\left\langle \xi\left(  t\right)  \xi\left(  t+s\right)
\right\rangle =\delta\left(  s\right)  $, and $\sigma=2\mu k_{b}T^{^{\prime}}$
with $k_{b}$ and $T^{^{\prime}}$ being the Boltzmann constant and temperature,
respectively. Also, we assume without loss of generality that $f\left(
0\leqslant t\leqslant T^{\ast}\right)  \geqslant0$ and redefine here the
impulse transmitted by $f(t)$ (per unit of amplitude) as%
\begin{equation}
I\equiv\int_{nT}^{nT+T^{\ast}}f\left(  t\right)  dt>0,\ n=0,1,2,...\ .
\tag{A2}%
\end{equation}
Equation~(A1) has the associated energy equation
\begin{equation}
\frac{dE}{dt}=-\mu\overset{.}{x}^{2}+\gamma\overset{.}{x}f\left(  t\right)
+\sqrt{\sigma}\overset{.}{x}\xi\left(  t\right)  , \tag{A3}%
\end{equation}
where $E(t)\equiv\left(  m/2\right)  \overset{.}{x}^{2}\left(  t\right)
+U\left[  x\left(  t\right)  \right]  $ is the energy function. Integration of
Eq.~(A3) over the intervals $\left[  nT,nT+T^{\ast}\right]  $ and $\left[
nT+T^{\ast},\left(  n+1\right)  T\right]  $, $n=0,1,2,...$, yields
\begin{align}
E\left(  nT+T^{\ast}\right)   &  =E(nT)-\mu\int_{nT}^{nT+T^{\ast}}\overset
{.}{x}^{2}\left(  t\right)  dt\nonumber\\
&  +\sqrt{\sigma}\int_{nT}^{nT+T^{\ast}}\overset{.}{x}\left(  t\right)
\xi\left(  t\right)  dt\nonumber\\
&  +\gamma\int_{nT}^{nT+T^{\ast}}\overset{.}{x}\left(  t\right)  f\left(
t\right)  dt,\tag{A4}\\
E\left[  \left(  n+1\right)  T\right]   &  =E(nT+T^{\ast})-\mu\int
_{nT+T^{\ast}}^{(n+1)T}\overset{.}{x}^{2}\left(  t\right)  dt\nonumber\\
&  +\sqrt{\sigma}\int_{nT+T^{\ast}}^{(n+1)T}\overset{.}{x}\left(  t\right)
\xi\left(  t\right)  dt\nonumber\\
&  +\gamma\int_{nT+T^{\ast}}^{(n+1)T}\overset{.}{x}\left(  t\right)  f\left(
t\right)  dt, \tag{A5}%
\end{align}
respectively, where the second integrals in Eqs.~(A4) and (A5) are considered
in the Stratonovich sense. After applying the first mean value theorem for
integrals [24] to the last integrals on the right-hand sides of Eqs.~(A4) and
(A5), using Eq. (A2), and recalling that $f(t)$ is a zero-mean function, one
obtains
\begin{align}
E\left(  nT+T^{\ast}\right)   &  =E(nT)-\mu\int_{nT}^{nT+T^{\ast}}\overset
{.}{x}^{2}\left(  t\right)  dt\tag{A6}\\
&  +\sqrt{\sigma}\int_{nT}^{nT+T^{\ast}}\overset{.}{x}\left(  t\right)
\xi\left(  t\right)  dt+\gamma\overset{.}{x}_{n}I,\nonumber\\
E\left[  \left(  n+1\right)  T\right]   &  =E(nT+T^{\ast})-\mu\int
_{nT+T^{\ast}}^{(n+1)T}\overset{.}{x}^{2}\left(  t\right)  dt\tag{A7}\\
&  +\sqrt{\sigma}\int_{nT+T^{\ast}}^{(n+1)T}\overset{.}{x}\left(  t\right)
\xi\left(  t\right)  dt-\gamma\overset{.}{x}_{n}^{\prime}I,\nonumber
\end{align}
respectively, where the discrete variables $\overset{.}{x}_{n}\equiv
\overset{.}{x}\left(  t_{n}\right)  ,\overset{.}{x}_{n}^{\prime}\equiv
\overset{.}{x}\left(  t_{n}^{\prime}\right)  $, with $t_{n}$ and
$t_{n}^{\prime}$ being unknown instants which only have to satisfy the
respective relationships $nT\leqslant t_{n}\leqslant nT+T^{\ast}$ and
$nT+T^{\ast}\leqslant t_{n}^{\prime}\leqslant(n+1)T$, according to the first
mean value theorem for integrals. After adding Eqs.~(A6) and (A7) from $n=0$
to $n=N-1$ and dividing the result by $NT$, one obtains%
\begin{align}
\frac{E\left(  NT\right)  -E\left(  0\right)  }{NT}  &  =-\frac{\mu}{NT}%
\int_{0}^{NT}\overset{.}{x}^{2}\left(  t\right)  dt\nonumber\\
&  +\gamma I\sum_{n=0}^{N-1}\left[  \frac{\overset{.}{x}_{n}-\overset{.}%
{x}_{n}^{\prime}}{NT}\right] \nonumber\\
&  +\frac{\sqrt{\sigma}}{NT}\int_{0}^{NT}\overset{.}{x}\left(  t\right)
\xi\left(  t\right)  dt. \tag{A8}%
\end{align}
Upon taking the limit $N\rightarrow\infty$ in Eq.~(A8), averaging over
different realizations of noise, and recalling that the system (A1) is
dissipative and that $\xi\left(  t\right)  $ is a stationary random process
which cannot contain a shot noise component, one finally obtains%
\begin{equation}
\left\langle \left\langle \overset{.}{x}^{2}\right\rangle \right\rangle
=\frac{\gamma I}{\mu}\left[  \left\langle \left\langle \overset{.}{x}%
_{n}\right\rangle \right\rangle -\left\langle \left\langle \overset{.}{x}%
_{n}^{\prime}\right\rangle \right\rangle \right]  +\frac{\sqrt{\sigma}}{\mu
}\left\langle \left\langle \overset{.}{x}\xi\right\rangle \right\rangle .
\tag{A9}%
\end{equation}
The following remarks are now in order. First, $\left\langle \left\langle
\overset{.}{x}_{n}\right\rangle \right\rangle $ provides the average of the
particle's velocity when $\overset{.}{x}$ is measured exclusively at certain
instants for which $f(t)$ has the same sign as the acceleration $\overset
{..}{x}$ [cf. Eq.~(A1)], i.e., when $f(t)$ tends to yield an increase in the
particle's velocity, while $\left\langle \left\langle \overset{.}{x}%
_{n}^{\prime}\right\rangle \right\rangle $ does the same when $f(t)$ has the
opposite sign to $\overset{..}{x}$, i.e., when $f(t)$ tends to yield a
decrease in the particle's velocity. One sees from Eq.~(A9) that the effect of
the difference $\left\langle \left\langle \overset{.}{x}_{n}\right\rangle
\right\rangle -\left\langle \left\langle \overset{.}{x}_{n}^{\prime
}\right\rangle \right\rangle $ on the average kinetic energy per unit of mass
is modulated by the impulse per unit of amplitude, while keeping the remaining
parameters constant. Second, increasing the noise strength from $\sigma=0$
activates the term $\left\langle \left\langle \overset{.}{x}\xi\right\rangle
\right\rangle $, which can be positive or negative. Third, one has
$\lim_{m\rightarrow0}\left[  E(NT)-E(0)\right]  /(NT)=\left[
U(NT)-U(0)\right]  /(NT)$ and hence Eq. (A9) remains valid in the overdamped
limiting case.

\subsection{Complementary case of harmonic excitations}

\noindent Let us consider the case of harmonic excitations in Eq.~(2) when the
roles of the excitations $\eta\cos\left(  \omega t\right)  $ and $\left(
1-\eta\right)  \cos\left(  2\omega t+\varphi\right)  $ are interchanged, i.e.,
the Langevin equation now reads%
\begin{align}
\overset{.}{x}+\sin\left[  x-\gamma(1-\eta)\cos\left(  2\omega t+\varphi
\right)  \right]   &  =\sqrt{\sigma}\xi\left(  t\right) \nonumber\\
&  +\gamma\eta\cos\left(  \omega t\right)  . \tag{A10}%
\end{align}
In the reference frame associated with the vibrating potential, one then
obtains%
\begin{align}
\overset{.}{z}+\sin z  &  =\gamma\left[  \eta\cos\left(  \omega t\right)
+2\omega\left(  1-\eta\right)  \sin\left(  2\omega t+\varphi\right)  \right]
\nonumber\\
&  +\sqrt{\sigma}\xi\left(  t\right)  , \tag{A11}%
\end{align}
where $z(t)\equiv x(t)-\gamma(1-\eta)\cos\left(  2\omega t+\varphi\right)  $.
Once again, ratchet universality predicts that the optimal value of the
relative amplitude $\eta$ comes from the condition that the amplitude of
$\cos\left(  \omega t\right)  $ must be twice as large as that of $\sin\left(
2\omega t+\varphi\right)  $ in Eq.~(A11), while the optimal values of the
initial phase difference are $\varphi=\varphi_{opt}\equiv\left\{  \pi
/2,3\pi/2\right\}  $ [16]. It therefore predicts the existence of a
\textit{different} (with respect to the case considered above, cf. Eq.~(18))
frequency-dependent optimal value of $\eta$:%
\begin{equation}
\eta_{opt}\equiv\frac{4\omega}{1+4\omega}, \tag{A12}%
\end{equation}
and, equivalently, a different optimal frequency for each value of $\eta$:
\begin{equation}
\omega_{opt}\equiv\frac{\eta}{4\left(  1-\eta\right)  }. \tag{A13}%
\end{equation}
Numerical simulations (as shown in Fig. 6) confirmed this new prediction over
a wide range of frequencies.
\begin{figure}
\includegraphics[width=.48\textwidth]{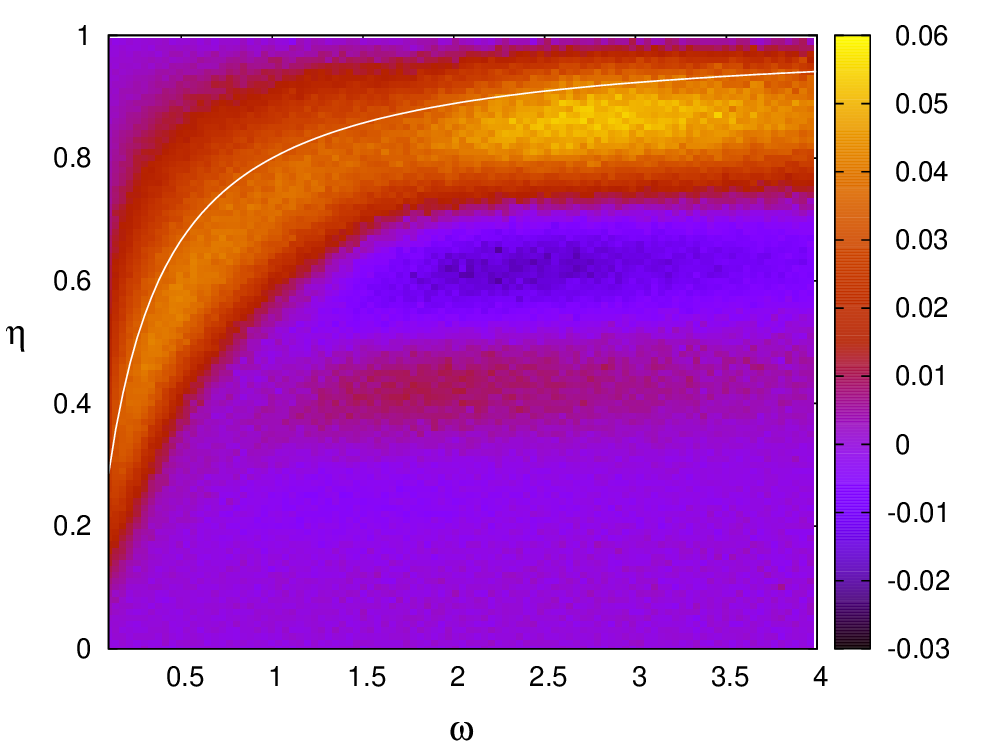}
\caption{Average velocity $\left\langle \left\langle \overset{.}{z}%
\right\rangle \right\rangle $ [cf. Eq.~(A11)] vs relative amplitude $\eta$ and
frequency $\omega$ for the parameters $\varphi=\varphi_{opt}=\pi
/2,\sigma=4,\gamma=8$. Also plotted (solid line) is the theoretical prediction
for the maximum average velocity [cf. Eq.~(A12)].}
\label{fig6}
\end{figure}

\subsection{Robustness against chaotic excitations}

\noindent In this subsection, we study the robustness of the universality
scenario against the presence of a bounded chaotic excitation instead of an
external periodic excitation. We shall consider the simple case $f(t)\equiv
\eta\cos\left(  \omega t+\varphi/2\right)  $, $g(t)\equiv\left(
1-\eta\right)  \alpha\overset{.}{y}(t),\omega\equiv2\pi/T,\eta\in\left[
0,1\right]  $ in Eq.~(2), i.e.,%
\begin{equation}
F(t)=F_{chaos}(t)\equiv\eta\omega\sin\left(  \omega t+\varphi/2\right)
+\left(  1-\eta\right)  \alpha\overset{.}{y}(t) \tag{A14}%
\end{equation}
in Eq.~(3), where $\overset{.}{y}(t)$ is a chaotic response of a master system
exhibiting the same underlying main frequency, $2\omega$, in its Fourier
spectrum [cf. Eq.~(17)], but cannot itself yield DRT. The value of $\alpha$ is
chosen in order for the excitations $\cos\left(  2\omega t+\varphi\right)  $
and $\alpha\overset{.}{y}(t)$ to have similar ranges. We considered the
following master system (damped driven pendulum)
\begin{equation}
\overset{..}{y}+K\sin y=-\delta\overset{.}{y}+F\cos\left(  2\omega
_{0}t\right)  , \tag{A15}%
\end{equation}
with the parameter values $\omega_{0}=0.5,K=2.25,\delta=0.375,F=2.48625,$ for
which the pendulum presents a chaotic attractor irrespective of the initial
conditions. Figure 7(a) shows the time series corresponding to the velocity
$\overset{.}{y}(t)$, and Fig.~7(b) shows the corresponding power spectrum
which presents its main peak at the frequency $2\omega_{0}$. Note the presence
of additional peaks at the frequencies $6\omega_{0},10\omega_{0},14\omega
_{0},...$, i.e., the underlying periodic solution, $f(t)$, only presents odd
harmonics and hence satisfies the shift symmetry $f(t+T/2)=-f(t)$ with
$T=\pi/\omega_{0}$. This means that the function $f(t)$ itself cannot yield
directed ratchet transport.
\begin{figure}
\includegraphics[width=.48\textwidth]{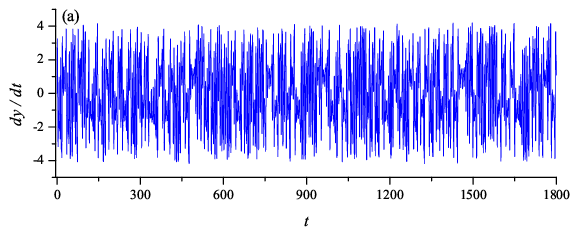}
\includegraphics[width=.48\textwidth]{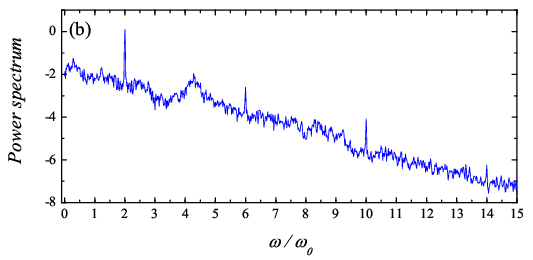}
\caption{(a) Velocity time series of $\overset{.}{y}(t)$, and (b) the
corresponding power spectrum ($\log_{10}\left\vert S\left(  \omega\right)
\right\vert $ versus $\omega/\omega_{0}$) associated with the damped driven
pendulum given by Eqs.~(A14) and (A15). Fixed parameters: $\omega
_{0}=0.5,K=2.25,\delta=0.375,F=2.48625$.}
\label{fig7}
\end{figure}

We found numerically the same dependence of the average velocity on $\eta$ as
in the biharmonic case [Eq.~(17)], but with a drastic decrease of the DRT
strength (see Fig.~8, top). Indeed, the presence of other noticeable harmonics
in the Fourier spectrum of $\overset{.}{y}(t)$ [cf. Fig. 7(b)] yields
interferences with the excitation $\eta\omega\sin\left(  \omega t\right)  $
which leads $F_{chaos}(t)$ to deviate from the optimal biharmonic
approximation [cf. Eq.~(10)]. This phenomenon and the inherent noise
background lead to $F_{chaos}(t)$ losing DRT effectiveness, but without
deactivating the DSB mechanism, and also to an additional decrease in the
optimal value of $\eta$ with respect to the corresponding deterministic
prediction [cf. Eq.~(18)]. This robustness is also manifest in the dependence
of the average velocity on $\varphi$ (see Fig. 8, bottom).
\begin{figure}
\includegraphics[width=.48\textwidth]{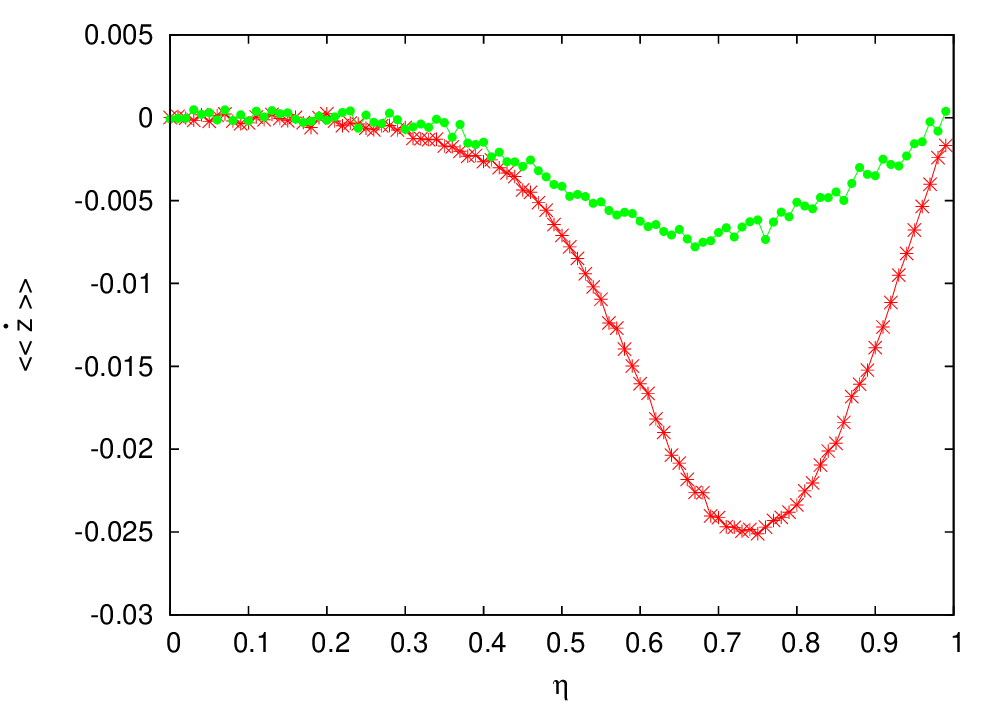}
\includegraphics[width=.48\textwidth]{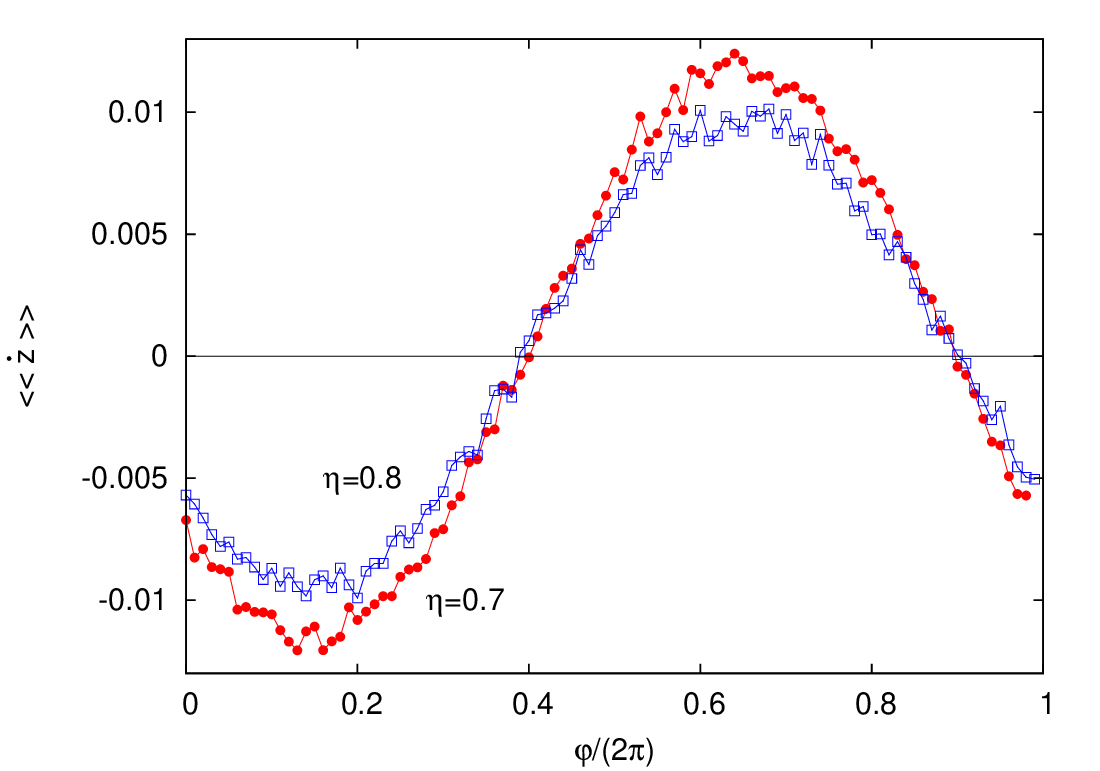}
\caption{Top: Average velocity $\left\langle \left\langle \overset{.}%
{z}\right\rangle \right\rangle $ [cf. Eq.~(3)] vs parameter $\eta$ for
$f(t)=\eta\cos\left(  \omega t+\varphi/2\right)  $, $\varphi=\varphi_{opt}=0$,
and two excitations $g(t)$ having the same underlying main frequency,
$2\omega$, in their Fourier spectrum: chaotic excitation [cf.~Eqs.~(A14) and
(A15); dots] and biharmonic excitation [cf.~Eq.~(17); stars]. Bottom:
$\left\langle \left\langle \overset{.}{z}\right\rangle \right\rangle $ vs
$\varphi$ for the chaotic excitation and $\eta=\left\{  0.7,0.8\right\}  $.
The lines connecting the symbols are solely to guide the eye. Fixed
parameters: $\gamma=8,T=4\pi,\sigma=5,\alpha=0.25$.}
\label{fig8}
\end{figure}


\begin{thebibliography}{99}                                                                                               %


\bibitem {1}R. P. Feynman, R. B. Leighton, and M. Sands, \textit{The Feynman
Lectures on Physics} (Addison Wesley, Reading, 1966) Vol. 1, Chapt. 46; J. M.
R. Parrondo and P. Espa\~{n}ol, Am. J. Phys. \textbf{64}, 1125 (1996).

\bibitem {2}P. Reimann, Phys. Rep. \textbf{361}, 57 (2002).

\bibitem {3}P. H\"{a}nggi and F. Marchesoni, Rev. Mod. Phys. \textbf{81}, 387 (2009).

\bibitem {4}F. J\"{u}licher, A. Ajdari, and J. Prost, Rev. Mod. Phys.
\textbf{69}, 1269 (1997).

\bibitem {5}M. Gu and C. M. Rice, Proc. Natl. Acad. Sci. U.S.A. \textbf{107},
521 (2010).

\bibitem {6}Jing-hui Li, Phys. Rev. E \textbf{67}, 061110 (2003).

\bibitem {7}S. Flach, O. Yevtushenko, and Y. Zolotaryuk, Phys. Rev. Lett.
\textbf{84}, 2358 (2000); S. Denisov, S. Flach, A. A. Ovchinnikov, O.
Yevtushenko, and Y. Zolotaryuk, Phys. Rev. E \textbf{66}, 041104 (2002).

\bibitem {8}A. B. Kolton, Phys. Rev. B \textbf{75}, 020201(R) (2007).

\bibitem {9}P. J. Mart\'{\i}nez and R. Chac\'{o}n, Phys. Rev. Lett.
\textbf{100}, 144101 (2008).

\bibitem {10}J. L. Mateos, Phys. Rev. Lett. \textbf{84}, 258 (2000); P.
Malgaretti, I. Pagonabarraga, and D. Frenkel, Phys. Rev. Lett. \textbf{109},
168104 (2011); A. Wickenbrock, D. Cubero, N. A. Abdul Wahab, P. Phoonthong,
and F. Renzoni, Phys. Rev. E \textbf{84}, 021127 (2011).

\bibitem {11}M. Rietmann, R. Carretero-Gonz\'{a}lez, and R. Chac\'{o}n, Phys.
Rev. A \textbf{83}, 053617 (2011).

\bibitem {12}G. Carapella and G. Costabile, Phys. Rev. Lett. \textbf{87},
077002 (2001); G. Carapella, Phys. Rev. B \textbf{63}, 054515 (2001).

\bibitem {13}F. R. Alatriste and J. L. Mateos, Physica A \textbf{372}, 263 (2006).

\bibitem {14}R. Gommers, S. Bergamini, and F. Renzoni, Phys. Rev. Lett.
\textbf{95}, 073003 (2005).

\bibitem {15}P. J. Mart\'{\i}nez and R. Chac\'{o}n, Phys. Rev. E \textbf{87},
062114 (2013); \textbf{88}, 019902(E) (2013); \textbf{88}, 066102 (2013).

\bibitem {16}R. Chac\'{o}n, J. Phys. A: Math. Theor. \textbf{40}, F413 (2007);
\textbf{43}, 322001 (2010).

\bibitem {17}M. Schiavoni, L. S\'{a}nchez-Palencia, F. Renzoni, and G.
Grynberg, Phys. Rev. Lett. \textbf{90}, 094101 (2003); R. Chac\'{o}n,
arXiv:1802.02826 (2018).

\bibitem {18}T. Salger \textit{et al}., Science \textbf{326}, 1241 (2009).

\bibitem {19}V. Berardi, J. Lydon, P. G. Kevrekidis, C. Daraio, and R.
Carretero-Gonz\'{a}lez, Phys. Rev. E \textbf{88}, 052202 (2013).

\bibitem {20}S. J. Lade, J. Phys. A: Math. Theor. \textbf{41}, 275103 (2008).

\bibitem {21}F. Marchesoni, Phys. Lett. A \textbf{119}, 221 (1986).

\bibitem {22}N. R. Quintero, J. A. Cuesta, and R. Alvarez-Nodarse, Phys. Rev.
E \textbf{81}, 030102(R) (2010); J. A. Cuesta, N. R. Quintero, and R.
Alvarez-Nodarse, Phys. Rev. X \textbf{3}, 041014 (2013).

\bibitem {23}S. Denisov, S. Flach, and P. H\"{a}nggi, Phys. Rep. \textbf{538},
77 (2014).

\bibitem {24}Gradshteyn, I. S. \& Ryzhik, I. M. \textit{Table of Integrals,
Series, and Products} (Academic Press, 1980).
\end{thebibliography}
\end{document}